\documentclass[twocolumn]{aastex631}

\usepackage{amsmath}
\usepackage{amssymb}

\makeatletter
\def\input@path{{./}{sections/}}
\makeatother
\graphicspath{{./}{figures/}}

\begin{document}
    % !TEX root = ../main.tex
\title{The LMT 2 Millimeter Receiver System (B4R). I. Overview and Results of Science Demonstration}

\author[0000-0002-8049-7525]{Ryohei Kawabe}
\affiliation{
    National Astronomical Observatory of Japan,
    National Institutes of Natural Sciences,
    2-21-1 Osawa,
    Mitaka,
    Tokyo 181-8588,
    Japan
}
\affiliation{
    Department of Astronomical Science,
    The Graduate University for Advanced Studies (SOKENDAI),
    2-21-1 Osawa,
    Mitaka,
    Tokyo 181-0015,
    Japan
}

\author[0000-0003-4521-7492]{Takeshi Sakai}
\affiliation{
    Graduate School of Informatics and Engineering,
    The University of Electro-Communications,
    1-5-1 Chofugaoka,
    Chofu,
    Tokyo 182-8585,
    Japan
}

\author[0000-0001-8153-1986]{Kunihiko Tanaka}
\affiliation{
    Department of Physics,
    Faculty of Science and Technology,
    Keio University,
    3-14-1 Hiyoshi,
    Yokohama,
    Kanagawa 223-8522,
    Japan
}

\author[0000-0002-9695-6183]{Akio Taniguchi}
\affiliation{
    Kitami Institute of Technology,
    165 Koen-cho,
    Kitami,
    Hokkaido 090-8507,
    Japan
}
\affiliation{
    Department of Physics,
    Graduate School of Science,
    Nagoya University,
    Furocho,
    Chikusa-ku,
    Nagoya,
    Aichi 464-8602,
    Japan
}

\author[0000-0001-6469-8725]{Bunyo Hatsukade}
\affiliation{
    National Astronomical Observatory of Japan,
    National Institutes of Natural Sciences,
    2-21-1 Osawa,
    Mitaka,
    Tokyo 181-8588,
    Japan
}
\affiliation{
    Department of Astronomical Science,
    The Graduate University for Advanced Studies (SOKENDAI),
    2-21-1 Osawa,
    Mitaka,
    Tokyo 181-0015,
    Japan
}

\affiliation{
    Institute of Astronomy,
    Graduate School of Science,
    The University of Tokyo,
    2-21-1 Osawa,
    Mitaka,
    Tokyo 181-0015,
    Japan
}

\author[0000-0003-4807-8117]{Yoichi Tamura}
\affiliation{
    Department of Physics,
    Graduate School of Science,
    Nagoya University,
    Furocho,
    Chikusa-ku,
    Nagoya,
    Aichi 464-8602,
    Japan
}

\author[0000-0002-1413-1963]{Yuki Yoshimura}
\affiliation{
    Institute of Astronomy,
    Graduate School of Science,
    The University of Tokyo,
    2-21-1 Osawa,
    Mitaka,
    Tokyo 181-0015,
    Japan
}

\author[0000-0002-4124-797X]{Tatsuya Takekoshi}
\affiliation{
    Kitami Institute of Technology,
    165 Koen-cho,
    Kitami,
    Hokkaido 090-8507,
    Japan
}
\affiliation{
    Institute of Astronomy,
    Graduate School of Science,
    The University of Tokyo,
    2-21-1 Osawa,
    Mitaka,
    Tokyo 181-0015,
    Japan
}

\author[0009-0005-5915-1035]{Tai Oshima}
\affiliation{
    National Astronomical Observatory of Japan,
    National Institutes of Natural Sciences,
    2-21-1 Osawa,
    Mitaka,
    Tokyo 181-8588,
    Japan
}
\affiliation{
    Department of Astronomical Science,
    The Graduate University for Advanced Studies (SOKENDAI),
    2-21-1 Osawa,
    Mitaka,
    Tokyo 181-0015,
    Japan
}

\author[0000-0001-8083-5814]{Masato Hagimoto}
\affiliation{
    Department of Physics,
    Graduate School of Science,
    Nagoya University,
    Furocho,
    Chikusa-ku,
    Nagoya,
    Aichi 464-8602,
    Japan
}

\author[0000-0003-2386-7427]{Teppei Yonetsu}
\affiliation{
    Department of Physics,
    Graduate School of Science,
    Osaka Metropolitan University,
    1-1 Gakuen-cho,
    Naka-ku,
    Sakai,
    Osaka 599-8531,
    Japan
}

\author[0000-0003-4402-6475]{Kotomi Taniguchi}
\affiliation{
    National Astronomical Observatory of Japan,
    National Institutes of Natural Sciences,
    2-21-1 Osawa,
    Mitaka,
    Tokyo 181-8588,
    Japan
}
\affiliation{
    Department of Astronomical Science,
    The Graduate University for Advanced Studies (SOKENDAI),
    2-21-1 Osawa,
    Mitaka,
    Tokyo 181-0015,
    Japan
}

\author[0000-0002-4052-2394]{Kotaro Kohno}
\affiliation{
    Institute of Astronomy,
    Graduate School of Science,
    The University of Tokyo,
    2-21-1 Osawa,
    Mitaka,
    Tokyo 181-0015,
    Japan
}
\affiliation{
    Research Center for the Early Universe,
    Graduate School of Science,
    The University of Tokyo,
    7-3-1 Hongo,
    Bunkyo-ku,
    Tokyo 113-0033,
    Japan
}

\author[0009-0007-6017-8395]{Hiroyuki Maezawa}
\affiliation{
    Department of Physics,
    Graduate School of Science,
    Osaka Metropolitan University,
    1-1 Gakuen-cho,
    Naka-ku,
    Sakai,
    Osaka 599-8531,
    Japan
}

\author{David H. Hughes}
\affiliation{
    Instituto Nacional de Astrof\'{i}sica,
    \'{O}ptica y Electr\'{o}nica,
    Luis Enrique Erro 1,
    Tonantzintla C.P. 72840,
    Puebla,
    M\'{e}xico
}

\author{Peter F. Schloerb}
\affiliation{
    Department of Astronomy,
    University of Massachusetts,
    Amherst,
    MA 01003,
    USA
}

\author[0000-0002-0758-3160]{Edgar Col\'{i}n-Beltr\'{a}n}
\affiliation{
    Instituto Nacional de Astrof\'{i}sica,
    \'{O}ptica y Electr\'{o}nica,
    Luis Enrique Erro 1,
    Tonantzintla C.P. 72840,
    Puebla,
    M\'{e}xico
}
\affiliation{
    Consejo Nacional de Ciencia y Tecnolog\'{i}a,
    Av. Insurgentes Sur 1582,
    Col. Cr\'{e}dito Constructor,
    Demarcaci\'{o}n Territorial Benito Ju\'{a}rez C.P. 03940,
    Ciudad de M\'{e}xico,
    M\'{e}xico
}

\author{Miguel Ch\'{a}vez Dagostino}
\affiliation{
    Instituto Nacional de Astrof\'{i}sica,
    \'{O}ptica y Electr\'{o}nica,
    Luis Enrique Erro 1,
    Tonantzintla C.P. 72840,
    Puebla,
    M\'{e}xico
}

\author[0009-0003-9025-6121]{V\'{i}ctor G\'{o}mez-Rivera}
\affiliation{
    Instituto Nacional de Astrof\'{i}sica,
    \'{O}ptica y Electr\'{o}nica,
    Luis Enrique Erro 1,
    Tonantzintla C.P. 72840,
    Puebla,
    M\'{e}xico
}
\affiliation{
    Corporaci\'{o}n Mexicana de Investigaci\'{o}n en Materiales S.A. de C.V.,
    M\'{e}xico
}

\author[0000-0001-9395-1670]{Arturo I. G\'{o}mez-Ruiz}
\affiliation{
    Instituto Nacional de Astrof\'{i}sica,
    \'{O}ptica y Electr\'{o}nica,
    Luis Enrique Erro 1,
    Tonantzintla C.P. 72840,
    Puebla,
    M\'{e}xico
}
\affiliation{
    Consejo Nacional de Ciencia y Tecnolog\'{i}a,
    Av. Insurgentes Sur 1582,
    Col. Cr\'{e}dito Constructor,
    Demarcaci\'{o}n Territorial Benito Ju\'{a}rez C.P. 03940,
    Ciudad de M\'{e}xico,
    M\'{e}xico
}

\author[0000-0002-4723-6569]{Gopal Narayanan}
\affiliation{
    Department of Astronomy,
    University of Massachusetts,
    Amherst,
    MA 01003,
    USA
}

\author{Iv\'{a}n Rodr\'{i}guez-Montoya}
\affiliation{
    Instituto Nacional de Astrof\'{i}sica,
    \'{O}ptica y Electr\'{o}nica,
    Luis Enrique Erro 1,
    Tonantzintla C.P. 72840,
    Puebla,
    M\'{e}xico
}
\affiliation{
    Consejo Nacional de Ciencia y Tecnolog\'{i}a,
    Av. Insurgentes Sur 1582,
    Col. Cr\'{e}dito Constructor,
    Demarcaci\'{o}n Territorial Benito Ju\'{a}rez C.P. 03940,
    Ciudad de M\'{e}xico,
    M\'{e}xico
}

\author[0000-0002-7344-9920]{David S\'{a}nchez-Arg\"{u}elles}
\affiliation{
    Instituto Nacional de Astrof\'{i}sica,
    \'{O}ptica y Electr\'{o}nica,
    Luis Enrique Erro 1,
    Tonantzintla C.P. 72840,
    Puebla,
    M\'{e}xico
}
\affiliation{
    Consejo Nacional de Ciencia y Tecnolog\'{i}a,
    Av. Insurgentes Sur 1582,
    Col. Cr\'{e}dito Constructor,
    Demarcaci\'{o}n Territorial Benito Ju\'{a}rez C.P. 03940,
    Ciudad de M\'{e}xico,
    M\'{e}xico
}

\author[0000-0001-9368-3143]{Yoshito Shimajiri}
\affiliation{
    Kyushu Kyoritsu University,
    1-8 Jiyugaoka,
    Yahatanishi-ku,
    Kitakyushu,
    Fukuoka,
    Fukuoka 807-8585,
    Japan
}
\affiliation{
    National Astronomical Observatory of Japan,
    National Institutes of Natural Sciences,
    2-21-1 Osawa,
    Mitaka,
    Tokyo 181-8588,
    Japan
}

\author[0000-0001-7915-5272]{Kamal Souccar}
\affiliation{
    Department of Astronomy,
    University of Massachusetts,
    Amherst,
    MA 01003,
    USA
}

\author[0000-0001-7095-7543]{Min S. Yun}
\affiliation{
    Department of Astronomy,
    University of Massachusetts,
    Amherst,
    MA 01003,
    USA
}

\author[0000-0002-5268-2221]{Tom J. L. C. Bakx}
\affiliation{
    Department of Space, Earth and Environment, Chalmers University of Technology,
    Gothenburg SE-412 96,
    Sweden
}
\affiliation{
    Department of Physics,
    Graduate School of Science,
    Nagoya University,
    Furocho,
    Chikusa-ku,
    Nagoya,
    Aichi 464-8602,
    Japan
}
\affiliation{
    National Astronomical Observatory of Japan,
    National Institutes of Natural Sciences,
    2-21-1 Osawa,
    Mitaka,
    Tokyo 181-8588,
    Japan
}

\author[0000-0001-5429-5762]{Kevin C. Harrington}
\affiliation{
    Joint ALMA Observatory,
    Alonso de C{\'o}rdova 3107,
    Vitacura,
    Santiago,
    Chile
}
\affiliation{
    National Astronomical Observatory of Japan,
    Los Abedules 3085 Oficina 701,
    Vitacura 763 0414,
    Santiago,
    Chile
}
\affiliation{
    European Southern Observatory,
    Alonso de C\'{o}rdova 3107,
    Vitacura,
    Casilla 19001,
    Santiago,
    Chile
}

\author[0000-0002-6375-7065]{Shinji Fujita}
\affiliation{
    The Institute of Statistical Mathematics,
    10-3 Midori-cho,
    Tachikawa,
    Tokyo,
    190-8562,
    Japan
}

\author[0000-0001-5431-2294]{Fumitaka Nakamura}
\affiliation{
    National Astronomical Observatory of Japan,
    National Institutes of Natural Sciences,
    2-21-1 Osawa,
    Mitaka,
    Tokyo 181-8588,
    Japan
}
\affiliation{
    Graduate School of Science,
    The University of Tokyo,
    7-3-1 Hongo,
    Bunkyo-ku,
    Tokyo 113-0033,
    Japan
}
\affiliation{
    Department of Astronomical Science,
    The Graduate University for Advanced Studies (SOKENDAI),
    2-21-1 Osawa,
    Mitaka,
    Tokyo 181-0015,
    Japan
}

\author[0000-0003-1054-4637]{O. S. Rojas-Garc\'{i}a}
\affiliation{
    Instituto Nacional de Astrof\'{i}sica,
    \'{O}ptica y Electr\'{o}nica,
    Luis Enrique Erro 1,
    Tonantzintla C.P. 72840,
    Puebla,
    M\'{e}xico
}

\author{Alfredo A. Monta\~{n}a Barbano}
\affiliation{
    Instituto Nacional de Astrof\'{i}sica,
    \'{O}ptica y Electr\'{o}nica,
    Luis Enrique Erro 1,
    Tonantzintla C.P. 72840,
    Puebla,
    M\'{e}xico
}

\author[0000-0001-8216-9800]{Javier Zaragoza-Cardiel}
\affiliation{
    Instituto Nacional de Astrof\'{i}sica,
    \'{O}ptica y Electr\'{o}nica,
    Luis Enrique Erro 1,
    Tonantzintla C.P. 72840,
    Puebla,
    M\'{e}xico
}
\affiliation{
    Consejo Nacional de Ciencia y Tecnolog\'{i}a,
    Av. Insurgentes Sur 1582,
    Col. Cr\'{e}dito Constructor,
    Demarcaci\'{o}n Territorial Benito Ju\'{a}rez C.P. 03940,
    Ciudad de M\'{e}xico,
    M\'{e}xico
}

\correspondingauthor{Ryohei Kawabe}
\email{ryo.kawabe32@gmail.com}
\shortauthors{Kawabe et al.}
\shorttitle{LMT 2\,mm Receiver B4R}

\submitjournal{AJ}
\received{December 31, 2024}
\revised{May 9, 2025}
\accepted{October 12, 2025}

\keywords{
    Millimeter astronomy (1061)
    ---
    Heterodyne receivers (727)
    ---
    Spectroscopy (1558)
    ---
    Astrochemistry (75)
    ---
    High-redshift galaxies (734)
}

\begin{abstract}
    We report on the results of the on-sky test and science demonstration conducted with the 2\,mm receiver system, B4R, on the 50\,m Large Millimeter Telescope (LMT), located at an altitude of 4600\,m in Mexico.
    The B4R receiver was developed based on the dual-polarization sideband-separating mixer technology of the Atacama Large Millimeter/submillimeter Array, and is equipped with a fast Fourier transform digital spectrometer, XFFTS.
    The primary science objective is the spectroscopic redshift identification of high-redshift dusty star-forming galaxies, complementing the existing 3\,mm Redshift Search Receiver by enabling the detection of multiple carbon monoxide lines.
    Additionally, the B4R receiver broadens the range of science cases possible with the LMT, including astrochemistry, as the 2\,mm band encompasses unique molecular lines such as deuterated molecules and shock tracers.
    During on-site commissioning in 2018 and 2019, we successfully demonstrated on-the-fly mapping and position-switching observations toward the Orion Molecular Cloud 1 and bright high-redshift dusty star-forming galaxies, respectively.
    We confirmed that the installed B4R system largely met its basic performance specifications.
    Furthermore, we measured the LMT's aperture efficiencies across the entire B4R frequency range (130--160\,GHz), finding them to be roughly consistent with expectations based on a surface accuracy of 100\,$\mu$m  and the receiver optics design.
    These results with the B4R will enable the most sensitive single-dish spectroscopic observations at 2\,mm using the LMT.
\end{abstract}

    % !TEX root = ../main.tex
\section{Introduction}
\label{sec:introduction}

The 2\,mm (150\,GHz) range is rich in scientific opportunities, and the Band 4 receivers, covering an observing frequency range of 125--163\,GHz, were developed and installed on the Atacama Large Millimeter/submillimeter Array (ALMA) 12\,m array and the Atacama Compact Array \citep{asayama2014}.
One of the major scientific cases for ALMA Band 4 is the detection of redshifted carbon monoxide (CO) lines in dusty star-forming galaxies (DSFGs), also known as submillimeter galaxies (SMGs).
Both the cosmic star formation rate density and the redshift distribution of SMGs have been found to peak around a redshift ($z$) of 2--3 and likely extend up to $z=6\text{--}7$ \citep[e.g.,][]{Madau2014}.
These dusty galaxies have been relatively easily surveyed through blank-field continuum observations at wavelengths of 0.3--2\,mm using ground-based telescopes such as the James Clerk Maxwell Telescope (e.g., \citealt{Smail1997, Hughes1998, Barger1998, Geach2017}), the Atacama Submillimeter Telescope Experiment (e.g., \citealt{tamura2009, Hatsukade2011, yun2012}), and the Herschel Space Observatory \citep[e.g.,][]{Eales2010}.
However, determining their spectroscopic redshifts in optical/infrared (opt/IR) observations remains challenging.
One approach to redshift determination with ALMA and other millimeter single-dish telescopes is to use CO ladders (i.e., $J=1\text{--}0$, 2--1, 3--2, 4--3; rotational transitions of CO) or the ionized carbon \textsc{[C\,ii]} line.
For the redshift range $z=2\text{--}4$, CO lines are the most useful, and even at higher redshifts ($z=4\text{--}6$), they can provide more precise redshifts than a single \textsc{[C\,ii]} line detection.
To determine an unambiguous redshift, at least two detections of CO lines (or a combination of CO and one or two neutral carbon \textsc{[C\,i]} lines) are required.
The 2\,mm receivers can supplement the second and third detections of CO lines in addition to the initial detection with 3\,mm receivers.
Thus, combining 2\,mm and 3\,mm receivers would be ideal for redshift determination(e.g., \citealt{Bakx2022}).

In addition, the 2\,mm range covers unique and important molecular lines for astrochemistry.
Deuterated molecular lines, DCO$^{+}$\,$J=2\text{--}1$ and DCN\,$J=2\text{--}1$, are examples observable in this range and allow us to estimate the D/H ratio, a key tool for investigating the chemical fractionation of deuterated molecules in cold star-forming molecular cores and commentary chemistry.
Additionally, major transitions of well-known shock tracers, such as methanol (CH$_{3}$OH), silicon monoxide (SiO $J=3\text{--}2$, $v=0$), and sulfur oxides (SO and SO$_{2}$), can be observed in this range.
Moreover, it covers various dense molecular gas tracers, including carbon monosulfide lines, such as CS\,$J=3\text{--}2$ and its isotopologues (C$^{34}$S, C$^{33}$S, $^{13}$CS), as well as formaldehyde (H$_{2}$CO) and cyanoacetylene (HC$_{3}$N) lines.

Here, we developed the 2\,mm receiver system, B4R, for the 50\,m Large Millimeter Telescope (LMT; \citealt{hughes2020}).
The primary goal of the B4R is to determine the spectroscopic redshift of dusty galaxies in conjunction with the 3\,mm receiver on the LMT, the Redshift Search Receiver (RSR; \citealt{Erickson2007}).
The LMT is the largest single-dish telescope capable of 2\,mm observations, and the addition of the 2\,mm receiver, equipped with a high spectral resolution spectrometer, significantly enhances its capabilities as a redshift search machine and expands its potential for a variety of scientific investigations.
In this paper, we provide an overview of the B4R system and present the results of recent science demonstration observations.

    % !TEX root = ../main.tex
\section{B4R System and Installation}
\label{sec:b4r-system-and-installation}

The B4R system consists of a single-beam, dual-polarization, sideband-separating (2SB), superconductor-insulator-superconductor (SIS) 2\,mm receiver and a fast Fourier transform (FFT) digital spectrometer, XFFTS.
In the following subsections, both the 2\,mm receiver and the digital spectrometer are described, along with details on their installation, sky conditions, and other relevant factors during the on-sky test and science demonstration.

\subsection{The 2\,mm Receiver}
\label{sec:the-2-mm-receiver}

\begin{figure}[htbp]
    \centering

    \includegraphics[width=\linewidth]{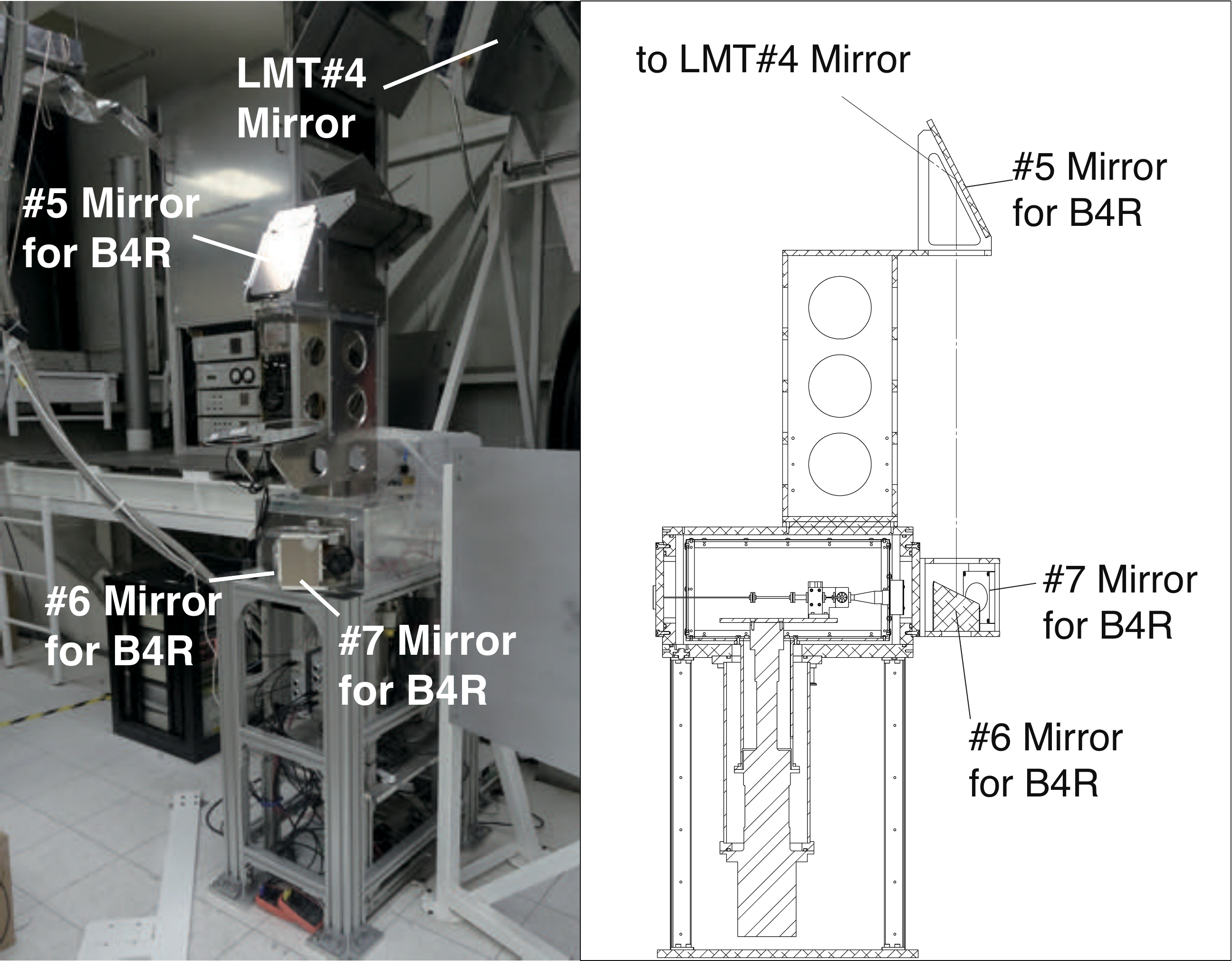}
    \caption {
        Left: photo of the B4R in the LMT Receiver Cabin.
        Right: drawing of the B4R and its warm optics.
    }
    \label{fig:b4r-photo}
\end{figure}

\begin{figure*}[htbp]
    \centering
    \includegraphics[width=\linewidth]{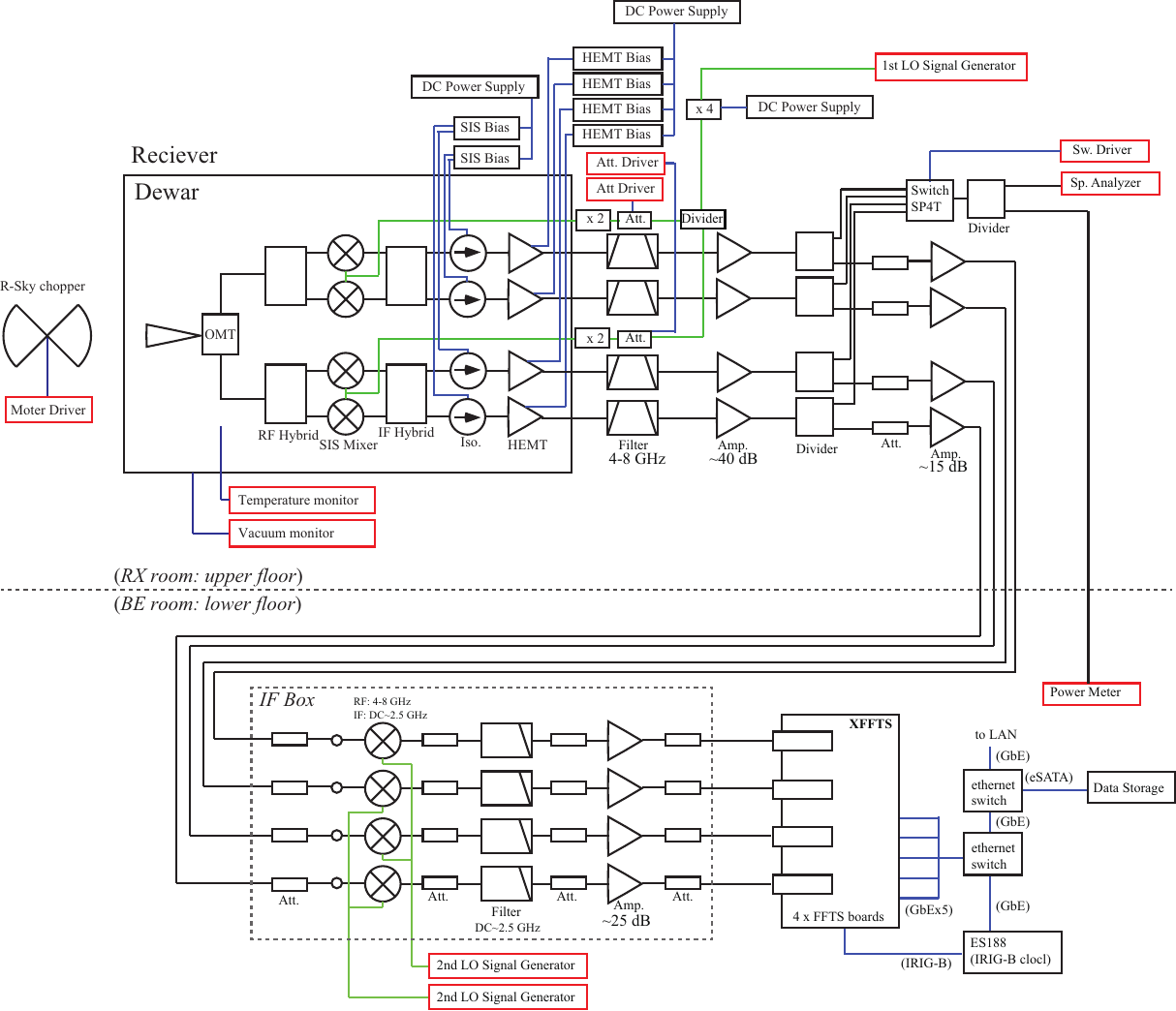}
    \caption {
        Block diagram of the B4R.
        The upper part is installed in the receiver cabin, and the lower part is located in the backend room.
        The components highlighted with red squares are controlled by computers.
    }
    \label{fig:block-diagram}
\end{figure*}

The 2\,mm receiver is based on the ALMA Band\,4 2SB mixer \citep{asayama2014}.
We redesigned the receiver dewar to accommodate a two-stage Sumitomo Gifford-McMahon 4\,K cryocooler (RDK-408D2P).
The optics of the B4R is designed to fully illuminate the 50\,m diameter of the LMT telescope with an edge taper of -12\,dB.
The warm optics layout for the B4R is shown in Figure~\ref{fig:b4r-photo}.
It includes a flat pick-up mirror (the \#5 mirror), followed by a flat mirror (\#6) and an ellipsoidal mirror (\#7).
The specifications of the receiver are summarized in Table~\ref{tab:spec}, along with those of the XFFTS spectrometer.
The first intermediate frequency (IF) range is 4--8\,GHz, and the first local oscillator (LO) consists of a synthesizer and a multiplier chain.
The first IF is down-converted using a second LO signal from another synthesizer to produce the final IF signal DC--2.5\,GHz.
Four sets of IF signals are fed into the spectrometer via bandpass filters.
During the on-sky test and science demonstration observations, the first IFs (5.6--8.1\,GHz) were down-converted to the baseband (DC--2.5\,GHz) using a second LO frequency of 8.1\,GHz.
The receiver block diagram is shown in Figure~\ref{fig:block-diagram}.

\begin{table}[htbp]
    \centering
    \caption{Specification of the B4R system}
    \label{tab:spec}
    \begin{tabular}{@{\hspace{1mm}}l@{\hspace{1mm}}l@{\hspace{1mm}}l}
        Item & Specification & Comment\\
        \hline
        RF frequency & 125--163\,GHz & ALMA Band\,4 spec.\\
        Polarization & Two, Linear & ALMA Band\,4 spec.\\
        Image rejection ratio & $>13$\,dB & ALMA Band\,4 spec.\\
        $\rm T_{RX}$ & $<60$\,K & ALMA Band\,4 spec.\\
        IF frequency & 4--8\,GHz & ALMA Band\,4 spec.\\
        Backend & 4 XFFTS boards & Expandable to 8\\
        Bandwidth & 2.5\,GHz/board & 10\,GHz in total\\
        Number of channels & 32,768\,/board & 131,072 in total\\
        Frequency resolution & 76.3\,kHz & 0.16\,km\,s$^{-1}$\\
        \hline
    \end{tabular}
\end{table}

\subsection{Digital Spectrometer}
\label{sec:digital-spectrometer}

The XFFTS \citep{klein2012} is an FFT digital spectrometer developed by Radiometer Physics GmbH.
We utilize four XFFTS boards with the standard configuration, each providing a DC--2.5\,GHz IF coverage and 32,768 spectral channels.
The FFT processing is performed using a flat-top window function.
At an observing frequency of 140\,GHz, the corresponding channel separation and spectral resolution are 0.163\,km\,s$^{-1}$.
The XFFTS board can achieve up to 25 times higher resolution than the standard mode by replacing the firmware, albeit at the cost of reduced simultaneous frequency coverage \citep{klein2012}.
The IF/LO system is designed to be upgradable to a full eight-board XFFTS configuration; however, currently, only four boards are installed for the on-sky test and science demonstration.

We operate the XFFTS system with a 10\,Hz sampling rate.
The spectral data are stored on a solid state drive using the vendor-provided software, \texttt{FFTS}, running on the control machine.
The data are recorded in raw float binary format and time-stamped by a network time protocol clock.
Data conversion to software-readable formats (the MeasurementSet version 2.0 for the Common Astronomy Software Applications package, CASA; \citealt{CASA2022}) and frequency conversion from the topocentric frame are performed using a dedicated data reduction pipeline, \texttt{b4r}\footnote{\url{https://github.com/b4r-dev/b4r}}, developed by our team.

\subsection{Installation, On-sky Test, and Science Demonstration}
\label{sec:installation--on-sky-test--and-science-}

The installation of the B4R system on the LMT was successfully completed in 2018 March.
The subsequent on-sky test began in 2018 June, as described in the next section, along with the technical checkout of the B4R receiver performance, receiver optics alignment, spectrometer performance, and other system components in the receiver cabin of the LMT.
The initial on-sky test aimed to verify whether the B4R system and observing software functioned properly.
This was done through spectroscopic observations of bright sources such as Orion-KL, and continuum observations of strong 3C sources and planets such as Uranus.
Through this initial test, we confirmed that the B4R receiver system is suitable for astronomical observations, although a few issues were identified and needed to be addressed.
A more detailed on-sky test commenced in 2018 October to evaluate system performance and assess the technical feasibility of observing modes such as position switching (PSW) and on-the-fly (OTF) mapping, along with science demonstration observations.
Additional on-sky tests were also conducted in 2024 May.
Observations were carried out at night, as is standard for the LMT operations with other instruments \citep{hughes2020}.
The observational parameters used during the on-sky tests and science demonstration are listed in Table~\ref{tab:obs-parameter}.

\begin{table}[htbp]
    \centering
    \caption{Observational Parameters}
    \label{tab:obs-parameter}
    \begin{tabular}{llll}
        Year, Month, Date & Frequency & $T_{\mathrm{SYS}}^{\dag}$ & $\tau_{225\,\mathrm{GHz}}^{\ddag}$\\
        & & (GHz) & (K)\\
        \hline
        2018 Oct 2--8 & 127--159 & 100--300 & 0.15--0.40\\
        2019 Nov 10--29 & 127--162 & 100--200 & 0.1--0.3\\
    \end{tabular}
    \\
    $^\dag$ The values of a single sideband.
    $^\ddag$ The values were measured by the 225\,GHz radiometer of the LMT \citep{Ferrusca2014}.
\end{table}

    % !TEX root = ../main.tex
\section{Results of On-sky Test}
\label{sec:results-of-on-sky-test}

\subsection{Beam Map}
\label{sec:beam-map}

\begin{figure}[htbp]
    \centering
    \includegraphics[width=\linewidth]{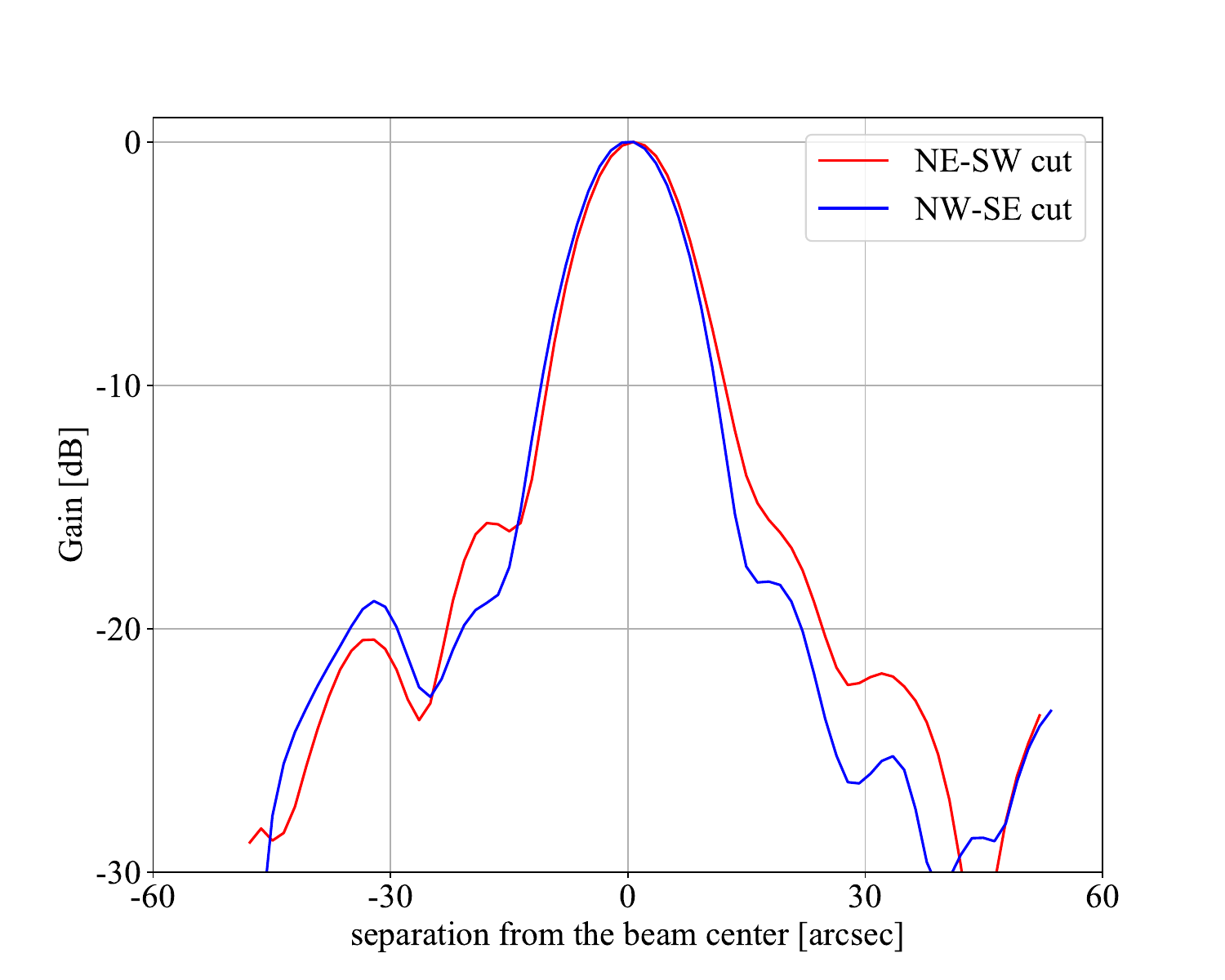}
    \caption{
        Beam pattern at 129.36\,GHz.
        The red and blue lines represent the radial profiles of the beam from northeast to southwest and from northwest to southeast in horizontal coordinates, respectively.
    }
    \label{fig:beam}
\end{figure}

The beam map at 129.36\,GHz was obtained by observing SiO masers ($J=3\text{--}2$, $v=1$) in Orion-KL in 2019 November.
Approximately 21 beam maps taken under good wind and optimal focus conditions were stacked to improve the signal-to-noise ratio (S/N).
The half power beamwidth (HPBW; $\theta_\mathrm{beam}$) was measured to be $\approx11''$ along two orthogonal directions, as shown in Figure~\ref{fig:beam}.
This measurement is consistent with the value calculated from the illumination pattern with an edge taper of -12\,dB, such as $1.2\times\lambda/D$, where $\lambda$ is the observing wavelength and $D$ is the telescope diameter.
Faint first and second sidelobes can be seen in the map, with their levels remaining below $\sim3$\% of the main lobe peak.

\subsection{Aperture Efficiency}
\label{sec:aperture-efficiency}

\begin{figure}[htbp]
    \centering
    \includegraphics[width=\linewidth]{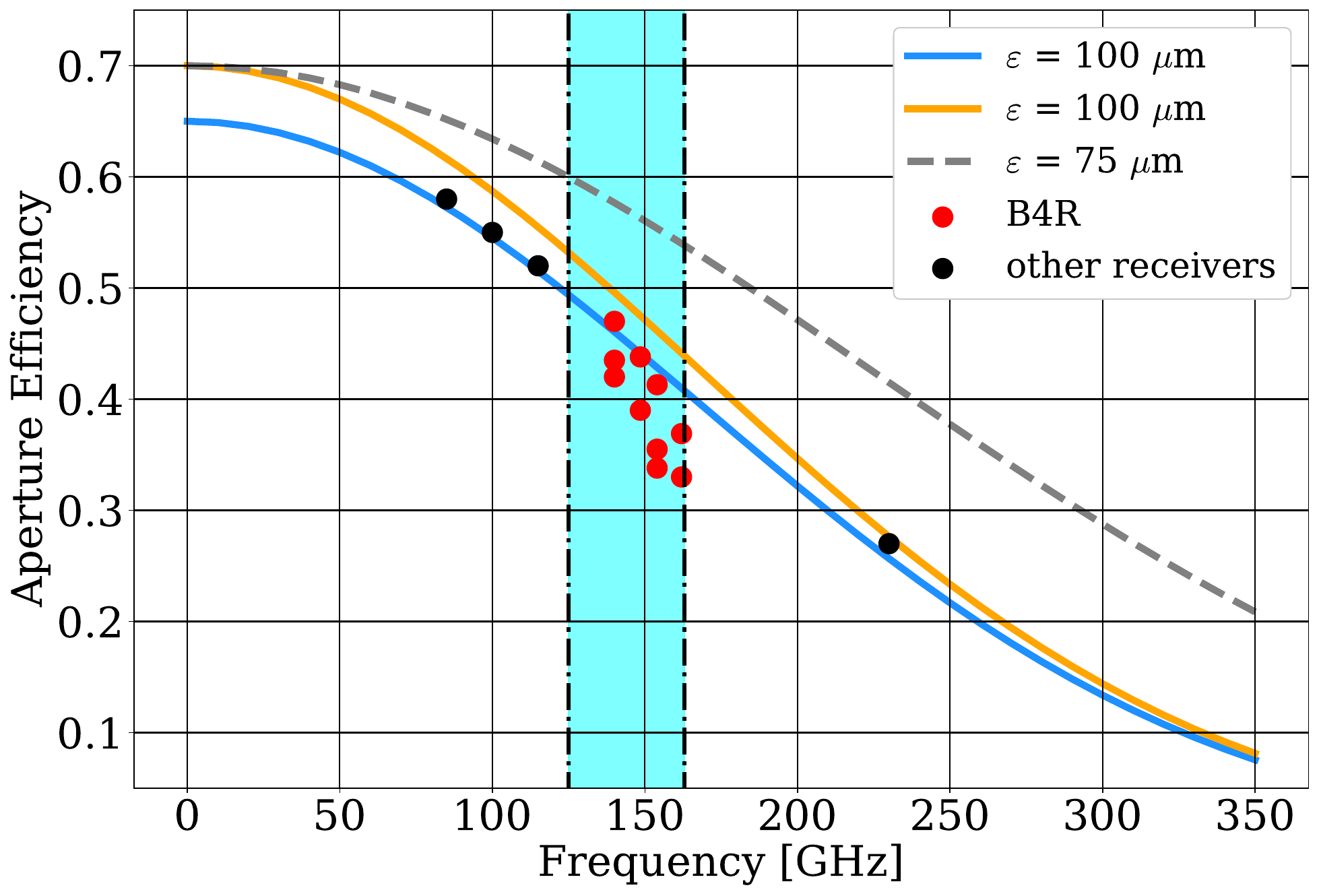}
    \caption {
        Aperture efficiency as a function of observing frequency.
        Red and black filled circles show the measured values with the B4R and other receivers, respectively.
        The blue line represents the case for the zero-frequency efficiency of 65\% and the rms surface error of 100\,$\mu$m.
        The other two lines (orange or gray) are for the rms surface error ($\varepsilon$) of 100 or 75\,$\mu$m and the zero-frequency efficiency of 70\%.
        Cyan fill shows the RF frequency range of the B4R.
    }
    \label{fig:eff}
\end{figure}

We observed Uranus in 2018 and 2019 at various frequencies to measure the aperture efficiency of the LMT.
The temperature scale was calibrated with the conventional chopper-wheel method.
In 2019, the measured efficiencies were 48\% to 33\% for 130 to 160\,GHz (see Figure~\ref{fig:eff}).
The systematic errors in the efficiency are estimated to be about 20\%, accounting for various uncertainties, including the assumed surface temperature of Uranus, the chopper-wheel calibration, and the measured beam sizes.
The obtained values roughly agree with those expected under the assumptions of a surface accuracy of 100\,$\mu$m and 65\% efficiency at zero frequency.
The zero-frequency efficiency is consistent with the value independently derived by multiplying two major factors: the illumination efficiency for the edge taper of -12\,dB, calculated to be approximately 77\% (e.g., using the \texttt{GRASP} software), and the blockage and shadowing efficiency, estimated to be roughly 85\% \citep{hughes2020}.
The efficiencies measured at frequencies above 150\,GHz are systematically lower than expected from the 100\,$\mu$m curve, and some values at lower frequencies also show unexpected reductions.
The degradation of pointing accuracy or focus during the efficiency measurements may have contributed to these lower efficiencies.
Therefore, we consider that it will be necessary to carefully reevaluate the efficiencies.
The beam parameters expected under the above assumptions (HPBW, aperture efficiency $\eta_{\mathrm{A}}$, main beam efficiency $\eta_{\mathrm{B}}$, and gain) are summarized in Table~\ref{tab:beam-parameter}.

\begin{table}[htbp]
    \centering
    \caption{Beam Parameters}
    \label{tab:beam-parameter}
    \begin{tabular}{lllll}
        Frequency & $\theta_{\mathrm{beam}}^{\dag}$ & $\eta_{\mathrm{A}}^{\ddag}$ & $\eta_{\mathrm{B}}^{\sharp}$ & Gain\\
        (GHz) & (arcsec) & (\%) & (\%) & (Jy\,K$^{-1}$)\\
        \hline
        130 & 10.97 & 48 & 57 & 2.91\\%0.34\\
        140 & 10.16 & 46 & 54 & 3.05\\%0.33\\
        150 & 9.51 & 44 & 52 & 3.21\\%0.31\\
        160 & 8.91 & 41 & 49 & 3.39\\%0.30\\
    \end{tabular}
    \\
    $^\dag$ Half power beamwidth.
    $^\ddag$ Aperture efficiency assuming the rms surface error of 100\,$\mu$m.
    $^\sharp$ Main beam efficiency.
\end{table}

\subsection{Pointing}
\label{sec:pointing}

The SiO\,$J=3\text{--}2$, $v=1$ maser emission was used for the ``offset'' pointing observations using the XFFTS spectrometer.
For this purpose, SiO maser sources with sufficient intensity were searched for near the target sources in advance.
Additionally, planets such as Mars and Uranus, as well as continuum sources like bright 3C sources, were also used.
Considering the accuracy of the global pointing derived from the 3\,mm observations using the multiple-beam receiver SEQUOIA, pointing sources with an apparent separation angle much smaller than 20$^{\circ}$--30$^{\circ}$ from the target seemed necessary.
Pointing calibrations were performed approximately every half to one hour, and the pointing drift between calibrations was typically measured to be 3$''$--5$''$ under stable weather conditions (e.g., wind speed $<10$\,m\,s$^{-1}$).

\subsection{Baseline Ripple Suppression}
\label{sec:baseline-ripple-suppression}

In the on-sky test conducted in 2018, we observed sinusoidal baseline ripples in the spectra, with a cycle of roughly 1.2\,GHz.
These ripples were particularly noticeable in the redshifted CO spectra of apparently hyperluminous infrared galaxies (HyLIRGs; $L_{\mathrm{IR}}>10^{13}\mathrm{L}_{\odot}$, e.g., \citealt{Fu2013, Ivison2013}), which are typically weak and wide in velocity (e.g., 500--1000 km\,s$^{-1}$ in full width at zero intensity, corresponding to 0.25--0.5\,GHz).
Such wide spectra would be challenging to detect.
We identified a possible root cause of the baseline ripples as a standing wave between the ZITEX (microporous polytetrafluoroethylene) film inserted as an IR filter at the 40\,K stage and the SIS mixer or feed horn at the 4\,K stage.
Initially, the ZITEX film was aligned perpendicularly to the feed horn axis.
Before the on-sky test in 2019, the ZITEX film was tilted by approximately 10$^{\circ}$ from its initial position.
As a result, the baseline ripples were reduced by a factor of more than 10 in amplitude, appearing to be less than about 1\,mK in $T_{\mathrm{a}}$ in the obtained CO spectra of high-$z$ HyLIRGs (see Figure~\ref{fig:high-z-spec}).

\subsection{Other Issues and Improvements}
\label{sec:other-issues-and-improvements}

During on-sky tests and science demonstration observations, we identified several issues with the B4R system.
Some of these issues were already resolved, while others were deferred as items for future upgrades.

The first issue was the vibration of the receiver's mechanical structure, which was induced by the acceleration and deceleration of the receiver cabin while the antenna was moving.
Before starting the test in 2019, an additional outer frame was installed to increase structural strength and reduce vibrations.
It was confirmed that the vibrations were reduced to a negligible level.

The second issue was that the two IF outputs from one of the two polarizations (A-pol) were slightly unstable compared to the other two outputs from the opposite polarization (B-pol).
These instabilities were observed during tests in both 2018 and 2019.
The possible causes investigated were an unstable mixer bias box for one SIS mixer in A-pol and an issue with the SIS mixer itself.
The bias box was replaced with a new one, and the electrical connection between the SIS junction and the bias line was also improved in 2024 September.
Following these modifications, A-pol appears to have recovered.
All results presented in Section~\ref{sec:results-of-science-demonstration} were obtained using B-pol.
If data from both A-pol and B-pol were available, the S/N in the maps or spectral lines would improve by a factor of $\sqrt{2}$.

The third issue concerns ``bad'' channels caused by spurious signals in the XFFTS spectra.
Several of the 32,768 spectral channels were identified as spurious due to the spectrometer itself.
These bad channels are primarily located at the first one or two channels, as well as at approximately 0.75, 1.25 (center), and 2.5\,GHz in the spectra (e.g., 0, 0.75, 1.25, and 2.5\,GHz).
The exact cause of these bad channels has not yet been determined.
Observations conducted in the topocentric reference frame appear to produce fixed bad channels, but the conversion to $v_{\mathrm{LSR}}$ can shift their locations and spread them across multiple channels in the final spectra.
Some offline data processing will be required to remove these bad channels at the earliest stage of spectral data analysis.

The fourth issue involves aliased spectral lines identified at the spectral band edges.
Some of these aliased lines originate from the second down-conversion of relatively strong lines from 4 to 8\,GHz to the baseband (0--2.5\,GHz), as well as aliasing caused by the analog-to-digital conversion (ADC; e.g., a 2.6\,GHz signal is aliased to 2.4\,GHz via the ADC).
This issue arises because the antialiasing filters at the baseband are not perfect; the current filters attenuate signals by only $\sim4\text{--}5$\,dB at 2.6\,GHz.
Careful identification of weaker spectral lines is particularly necessary near the band edges (see also Section~\ref{sec:spectral-scan-of-ori-kl} and \citealt{Yonetsu2025}).

    % !TEX root = ../main.tex
\section{Results of Science Demonstration}
\label{sec:results-of-science-demonstration}

\subsection{PSW Observations of High-$z$ HyLIRGs}
\label{sec:psw-observations-of-high-z-smgs}

We carried out PSW observations of seven apparent HyLIRGs at redshift $z\sim 2\text{--}4$ in 2019 November.
Each observational scan consisted of 10\,s hot-load measurement and a total of 600\,s of on-source and off-source switching observations.
Our targets were originally identified by Planck (Planck Catalog of Compact Sources; \citealt{planck2014}), Herschel (Herschel Stripe 82 Survey; \citealt{Viero2014}), and WISE (``AllWISE'' Data Release; \citealt{Cutri2013}), and their spectroscopic redshifts were determined through CO emission lines in previous studies \citep{Canameras2015, Geach2015, Harrington2016, Canameras2018, Harrington2021}.

\begin{figure*}[htbp]
    \centering
    \includegraphics[width=\linewidth]{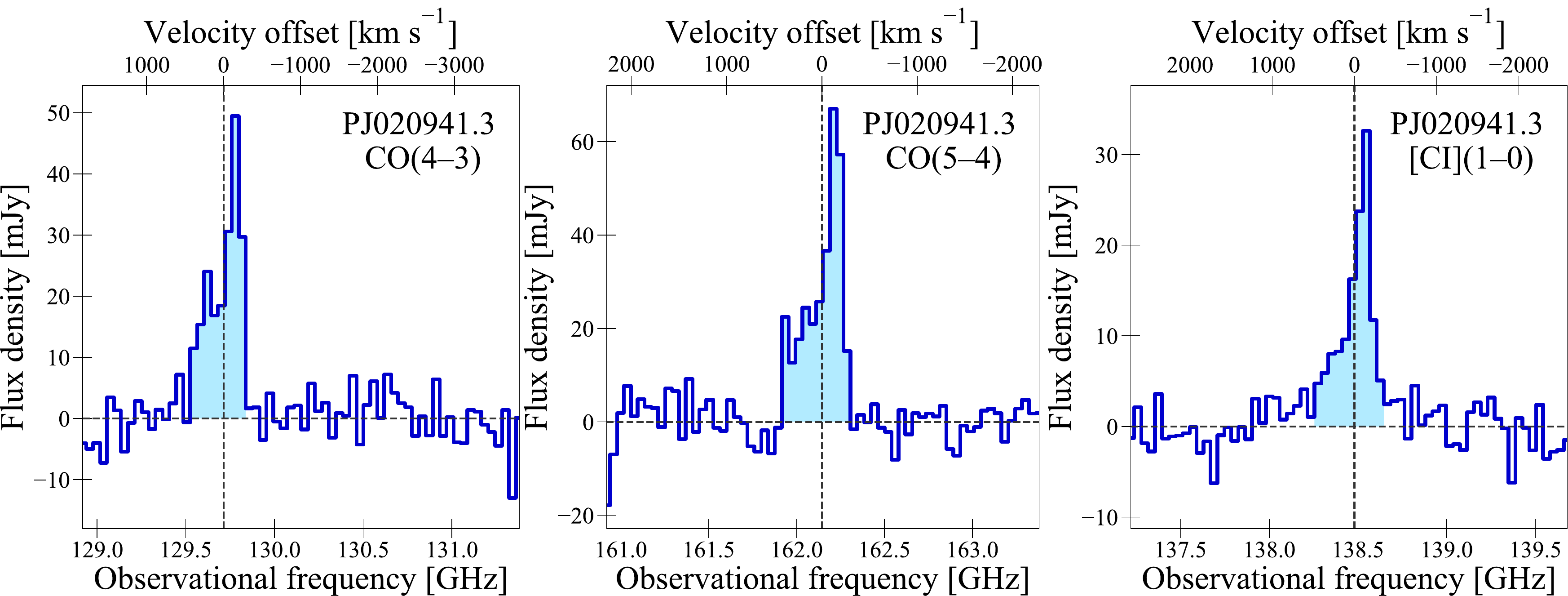}
    \caption{
        Redshifted CO and [\textsc{C\,i}] spectra covering a frequency range of 2.5\,GHz, taken toward one of the Planck-selected, strongly lensed HyLIRGs, PJ020941.3, at $z=2.55$.
        In each panel, $v=0$\,km\,s$^{-1}$ corresponds to the line redshift obtained from our observations (see Table~\ref{tab:high-z-observation}).
    }
    \label{fig:high-z-spec}
\end{figure*}

Here, we present three emission-line spectra of one of the targets, PJ020941.3, taken with the B4R, shown in Figure~\ref{fig:high-z-spec}.
The spectra cover a frequency range of 2.5\,GHz with a flat baseline.
The redshift of this target was determined to be $z_\mathrm{spec}=2.5534\pm0.0002$ \citep{Harrington2016} through the detection of the redshifted CO\,$J=3\text{--}2$ line ($\nu_\mathrm{obs} = 97.314$\,GHz) with the RSR installed on the LMT (see also \citealt{Geach2015}).
Subsequently, \citet{Harrington2021} reported multiple CO and \textsc{[C\,i]} line detections using IRAM and APEX and conducted detailed modeling of the gas's physical properties utilizing the comprehensive CO and \textsc{[C\,i]} data.

We observed three redshifted emission lines: CO\,$J=4\text{--}3$ ($\nu_\mathrm{obs} = 129.746$\,GHz), CO\,$J=5\text{--}4$ ($\nu_\mathrm{obs} = 162.174$\,GHz), and \textsc{[C\,i]}\,$^3\mathrm{P}_1\text{--}^3\mathrm{P}_0$ ($\nu_\mathrm{obs} = 138.504$\,GHz).
These emission lines were detected and reported in previous studies \citep{Geach2018, Harrington2019, Harrington2021}.
In Table~\ref{tab:high-z-observation}, we summarize the observation logs of PJ020941.3 obtained with the B4R mounted on the LMT.
All three lines detected in this source are presented in Figure~\ref{fig:high-z-spec} (all other spectra for line-detected sources will be presented in Table~\ref{tab:high-z-observation_appendix} and Figure~\ref{fig:high-z-observation_appendix} in Appendix~\ref{sec:spectra-of-other-emission-line-detected-in-high-z-smgs}.)

\begin{table*}[htbp]
    \centering
    \caption{Observation logs of one of the high-$z$ targets obtained with the B4R on the LMT}
    \label{tab:high-z-observation}
    \begin{tabular}{llll}
        \hline
        & CO\,$J=4\text{--}3$ & CO\,$J=5\text{--}4$ & \textsc{[C\,i]}\,$^3\mathrm{P}_1\text{--}^3\mathrm{P}_0$\\
        \hline\hline
        Target name & \multicolumn{3}{c}{PJ020941.3}\\
        Target coordinates (J2000) & \multicolumn{3}{c}{$\alpha=02^\mathrm{h}09^\mathrm{m}41.3^\mathrm{s}$, $\delta=+00^\circ15^\prime59''$} \\
        \hline
        Observation date & 2019 Nov 21, 22, 26 & 2019 Nov 26 & 2019 Nov 26, 27\\
        The total number of scans & 8 & 4 & 4 \\
        $T_\mathrm{atm}$ [K] & 275.4--277.2 & 277.2--277.3 & 276.6--276.8\\
        Opacity at 220\,GHz (min, max) & (0.15, 0.35) & (0.16, 0.19) & (0.10, 0.14)\\
        System noise temperature [K] (min, max) & (106, 152) & (119, 124) & (95, 129)\\
        \hline
        B4R first-LO frequency [GHz] & 137.0 & 155.3 & 145.3\\
        XFFTS frequency range [GHz] & 128.9--131.4 & 160.9--163.4 & 137.2--139.7\\
        512-ch binned channel width [GHz] & 0.04 (90\,km\,s$^{-1}$) & 0.04 (72\,km\,s$^{-1}$) & 0.04 (84\,km\,s$^{-1}$) \\
        \hline
        Integration time at on-source position [s] & 600 & 600 & 900\\
        Integration time at off-source position [s] & 600 & 600 & 900\\
        Baseline subtracted function & linear & linear & linear $+$ sine curve\\
        \hline
        Line redshift & $2.5543\pm0.0001$ & $2.5540\pm0.0001$ & $2.5538\pm0.0001$\\
        Noise level [mJy] & 3.84 & 4.63 & 2.43\\
        \hline
    \end{tabular}
\end{table*}

After the observations, we carried out data reduction as follows.
First, we applied chopper-wheel calibration to each 10\,s measurement of the hot load and to a total of 300\,s of on-source and off-source observations.
Here, this observation set is referred to as a ``scan''.
Second, we replaced spurious signals (or bad channels) with random values following the noise distribution of the observation.
Next, we integrated the high-quality scans with S/N greater than 3.5.
We binned 512 frequency channels to improve the S/N ($N_\mathrm{freq}=64$ after binning), resulting in a velocity resolution of $\sim 80$\,km\,s$^{-1}$ at 145\,GHz.
After that, we applied linear baseline subtraction to the CO\,$J=4\text{--}3$ and CO\,$J=5\text{--}4$ spectra.
For the \textsc{[C\,i]}\,$^3\mathrm{P}_1\text{--}^3\mathrm{P}_0$ spectrum, we subtracted the baseline by combining a sinusoidal curve and a linear line to remove the baseline ripple.
Finally, we converted antenna temperature to flux density by $\eta_\mathrm{A}$ in Section~\ref{sec:aperture-efficiency}, $\theta_\mathrm{beam}\approx10''$ (at $\approx140$\,GHz), and $S_\mathrm{CO}\,[\mathrm{Jy}]=1.42/\eta_\mathrm{A}\times T_\mathrm{A}\,[\mathrm{K}]$.

\begin{figure}[htbp]
    \centering
    \includegraphics[width=\linewidth]{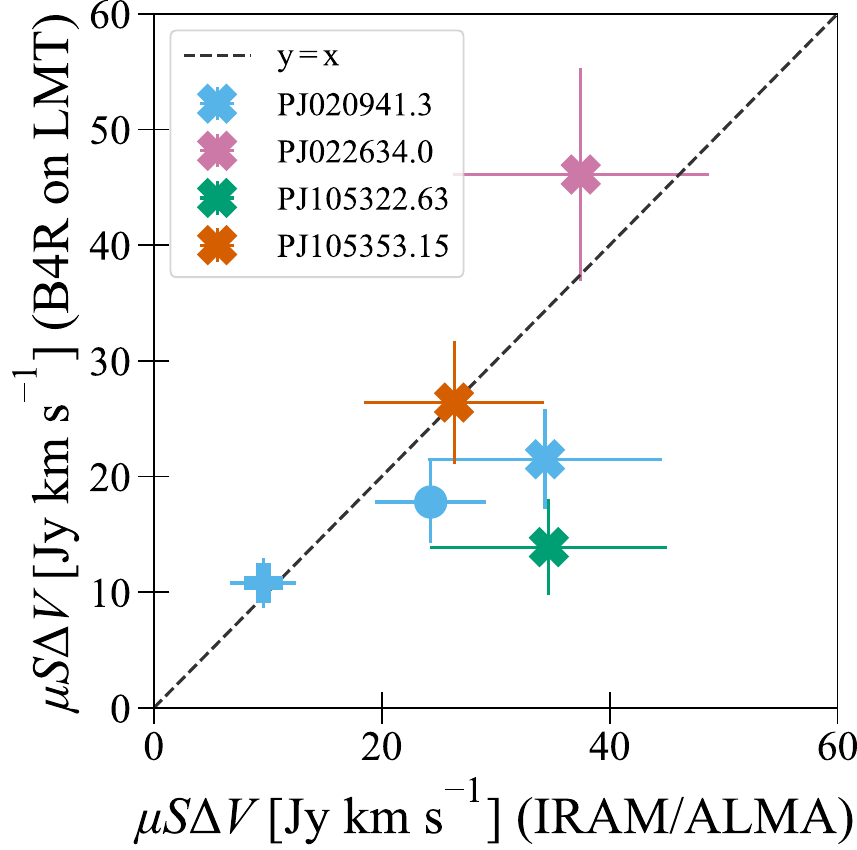}
    \caption{
        A comparison of the integrated CO or [\textsc{C\,i}] intensities between the B4R on the LMT and the IRAM 30\,m telescope or ALMA, as measured in \citet{Harrington2021}.
        The circle, crosses, and plus sign correspond to the intensities of CO\,$J=4\text{--}3$, CO\,$J=5\text{--}4$, and \textsc{[C\,i]}\,$^3\mathrm{P}_1\text{--}^3\mathrm{P}_0$, respectively.
        The different colors represent different targets.
        The systematic error in the intensity measured with the B4R is taken to be 20\% for each spectrum.
    }
    \label{fig:fluxcomparison}
\end{figure}

For the CO and [\textsc{C\,i}] spectra in four of the five line-detected sources (other than PJ011646.8) shown in Figures~\ref{fig:high-z-spec} and \ref{fig:high-z-spec_appendix}, we compared the spectral shapes and integrated intensities with those previously obtained by the IRAM 30\,m and ALMA \citep{Geach2018, Harrington2019, Harrington2021}.
As a result, we confirmed that the spectral shapes are consistent with each other.
To check the consistency between the integrated intensities from the B4R and those measured by IRAM 30\,m/ALMA in \citet{Harrington2021}, we compared them as shown in Figure~\ref{fig:fluxcomparison}.
Here, we adopt a systematic error of 20\% following \citet{Harrington2021}, which includes atmospheric and receiver instability, calibration of gain conversion $(\eta_\mathrm{A})$, baseline subtraction procedure, and pointing and focusing accuracy.
The line intensities obtained by the B4R are mostly consistent with those from IRAM 30\,m/ALMA.
However, we found an inconsistency for PJ105322.63, where the intensity from the B4R is $\sim40$\% of that from IRAM 30\,m.
The possible cause of this result is a large pointing error in the B4R observations due to the large angular distance of $\sim30$\,deg between the target and the pointing source.
In addition, this line was detected at the edge of the lower frequency side of the B4R, where the power response could have an issue.
(\citealt{Harrington2021} included an additional 5\%--10\% uncertainty for emission-line intensities detected at the lower frequency edge of the EMIR receiver on IRAM 30\,m, considering similar effects.)
In fact, the peak of our obtained spectrum is about twice fainter than that reported in \citet{Harrington2021}.

The CO\,$J = 4\text{--}3$ and CO\,$J = 5\text{--}4$ data were also analyzed using a new data-scientific method \citep{taniguchi2021}\footnote{Data analysis codes are available at \url{https://github.com/astropenguin/taniguchi-2021-analysis}.}, with only one scan for each CO line.
We confirmed that the spectral shapes obtained here are consistent with those presented in the paper.
The advantages and future prospects of these methods will be further discussed in Section~\ref{sec:data-scientific-methods-for-spectroscopic-observations}.

\subsection{OTF Mapping of OMC-1}
\label{sec:otf-mapping-of-omc-1}

\begin{figure}[htbp]
    \centering
    \includegraphics[width=\linewidth]{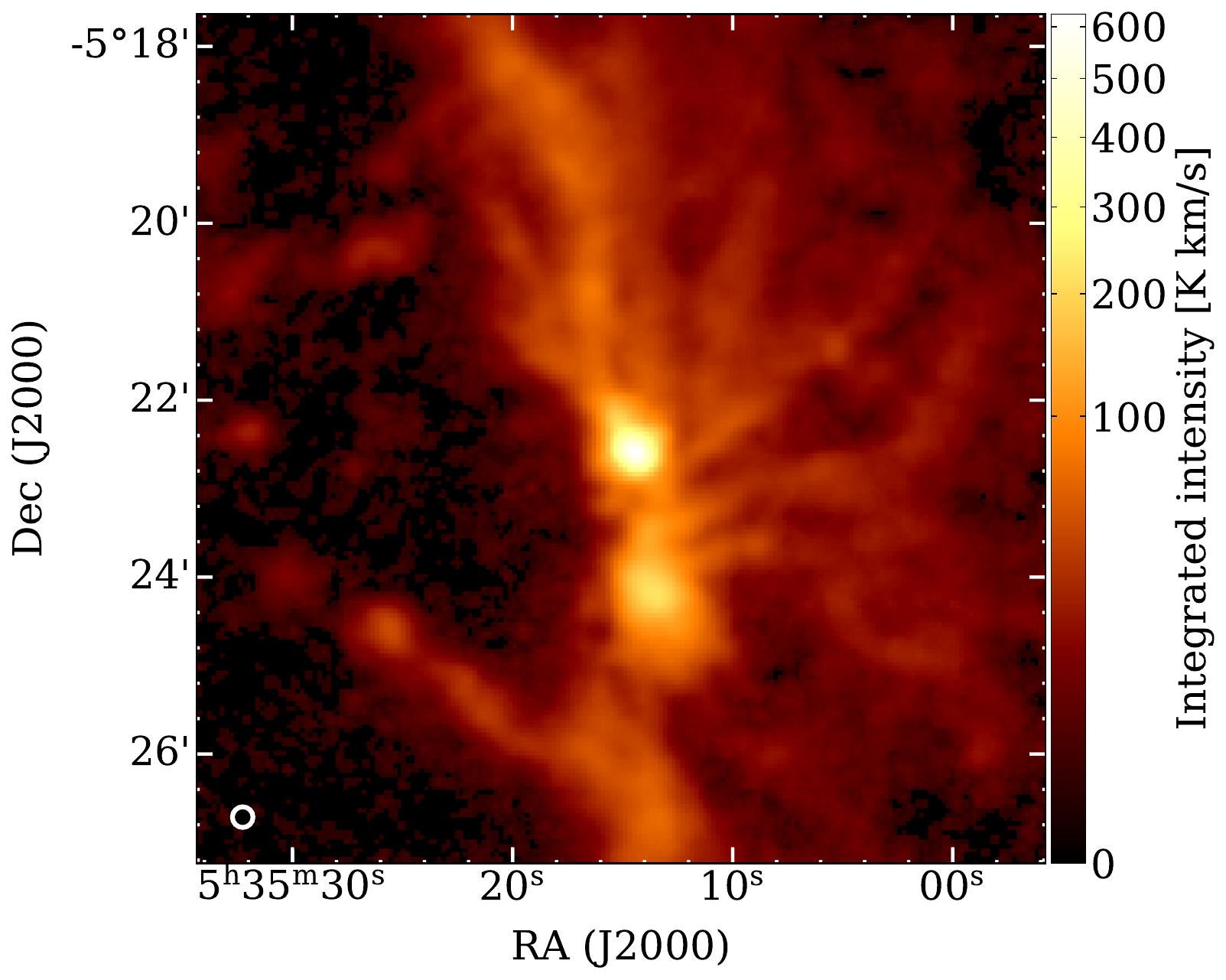}
    \caption {
        $10'\times10'$ integrated intensity map of CS\,$J=3\text{--}2$ (146.969033\,GHz) toward OMC-1 (including Orion-KL) observed in 2018.
    }
    \label{fig:otf1}
\end{figure}

\begin{figure*}[htbp]
    \centering
    \includegraphics[width=\linewidth]{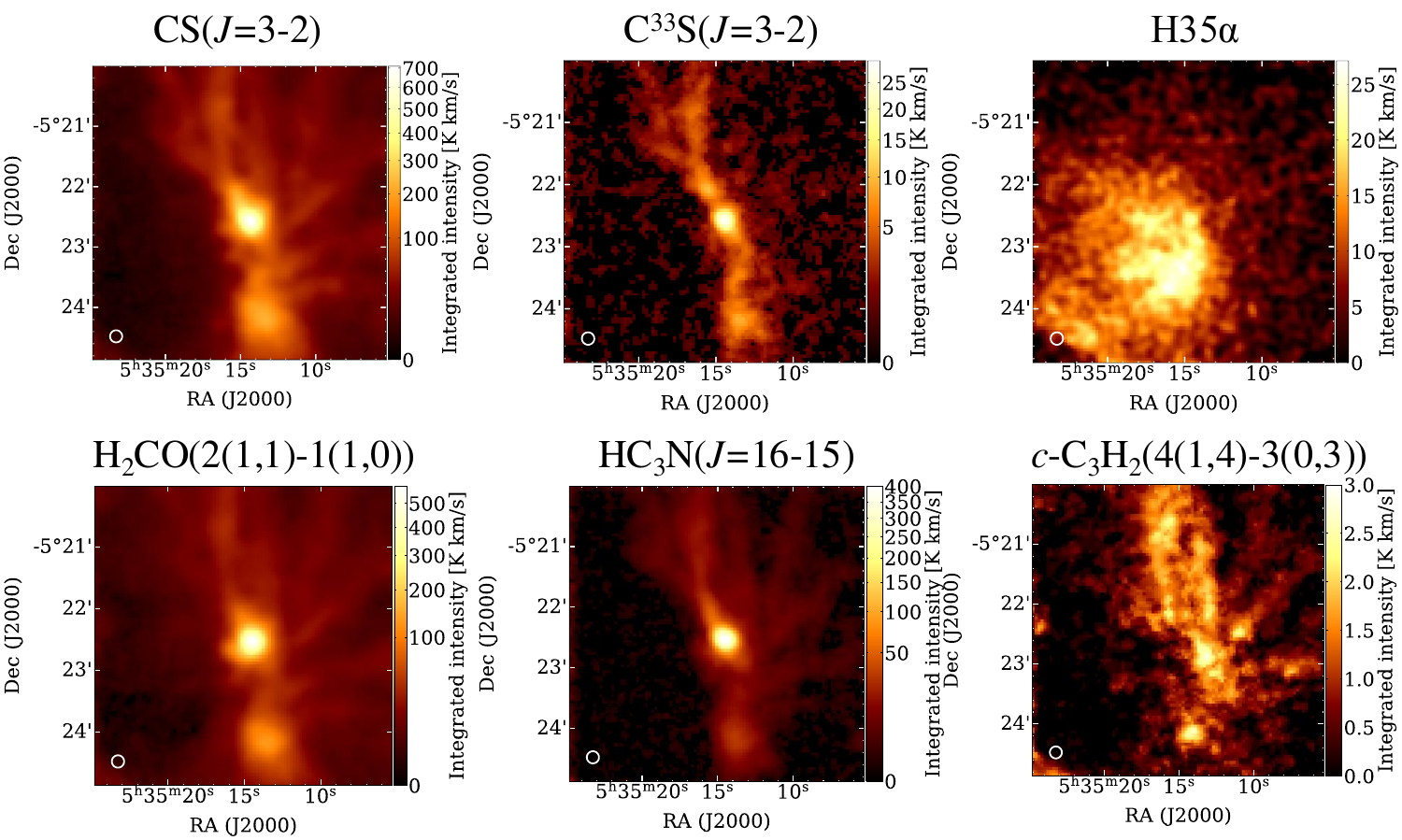}
    \caption{
        $5'\times5'$ integrated intensity maps of various emission lines toward OMC-1 observed in 2019.
        The rest frequencies and beam sizes are CS (146.969033\,GHz, $14''$), C$^{33}$S (145.7557316\,GHz, $14''$), H35$\alpha$ (147.046848\,GHz, $15''$), H$_2$CO (150.4983334\,GHz), HC$_3$N (145.560946\,GHz), \textit{c}-C$_3$H$_2$ (150.851908\,GHz), respectively.
    }
    \label{fig:otf2}
\end{figure*}

\begin{figure*}[htbp]
    \centering
    \includegraphics[width=\linewidth]{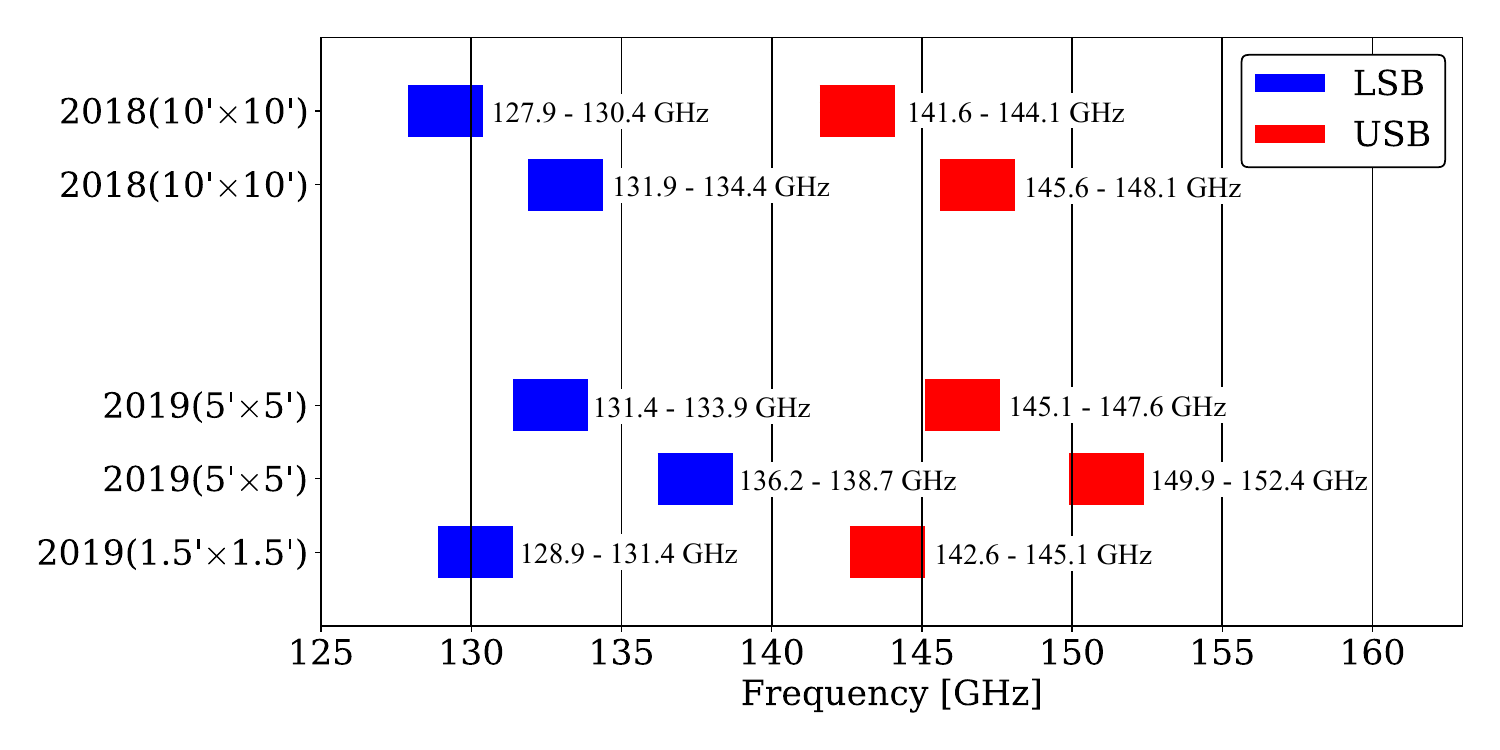}
    \caption {
        Observational frequency ranges for the five sets of OTF mapping toward OMC-1: two $10'\times10'$, two $5'\times5'$, and one $1.5'\times1.5'$ OTF mapping.
        The blue and red boxes represent the LSB and USB, respectively.
    }
    \label{fig:freq_band}
\end{figure*}

We carried out several OTF mapping observations of Orion Molecular Cloud 1 (OMC-1).
The OTF maps, with a size of $10'\times 10'$, were taken with two frequency settings in 2018, and the maps with $5'\times5'$ were taken with two other frequency settings in 2019.
Only x-scans along R.A. were performed in the OTF mapping, with scan separations of $7''$ and $3''$ in the 2018 and 2019 observations, respectively.
The total observing time for each map was 90 minutes for $10'\times10'$ maps and 75 minutes for $5'\times5'$ maps (the on-source time was 3030\,s and 2526\,s, respectively).
The OTF maps were made using a standard method with an appropriate gridding function, and 3D data cubes with 32,768 frequency channels in FITS format were created.
Data reduction and analysis, such as channel binning and making integrated intensity maps, were performed in CASA.
Examples of the OTF maps are shown in Figures~\ref{fig:otf1} and \ref{fig:otf2}, where no additional analysis, such as destriping, was applied.
Even simple observations and data reductions provide us with high-quality images of molecular lines.
The typical rms noise in channel maps after binning 8 or 10 frequency channels (corresponding to 1.8\,km\,s$^{-1}$) is $\sim0.25$\,K and $\sim0.1$\,K in $T_{\mathrm{a}}$ for the 2018 and 2019 OTF data, respectively.
The observed frequency coverages for both the lower sideband (LSB) and upper sideband (USB) are summarized in Figure~\ref{fig:freq_band}.
Note that the second LO frequency was set to 8.1\,GHz for all OTF mapping.

In 2019, $1.5'\times1.5'$ OTF maps were obtained from observations to make beam maps using SiO maser lines ($J=3\text{--}2$, $v=1$) as described in Section~\ref{sec:beam-map}.
The OTF maps of DCN and DCO$^+$\,$J=2\text{--}1$ lines were obtained using the beam map data, with no degradation of pointing accuracy due to wind during the observations \citep{Taniguchi2024-pt}.
Other deuterated molecules such as CCD and HDCO, CCS, and 2\,mm continuum emission were also successfully imaged using the beam map data.

\subsection{Spectral Scan of Orion-KL}
\label{sec:spectral-scan-of-ori-kl}

\begin{figure*}[htbp]
    \centering
    \includegraphics[width=0.95\linewidth]{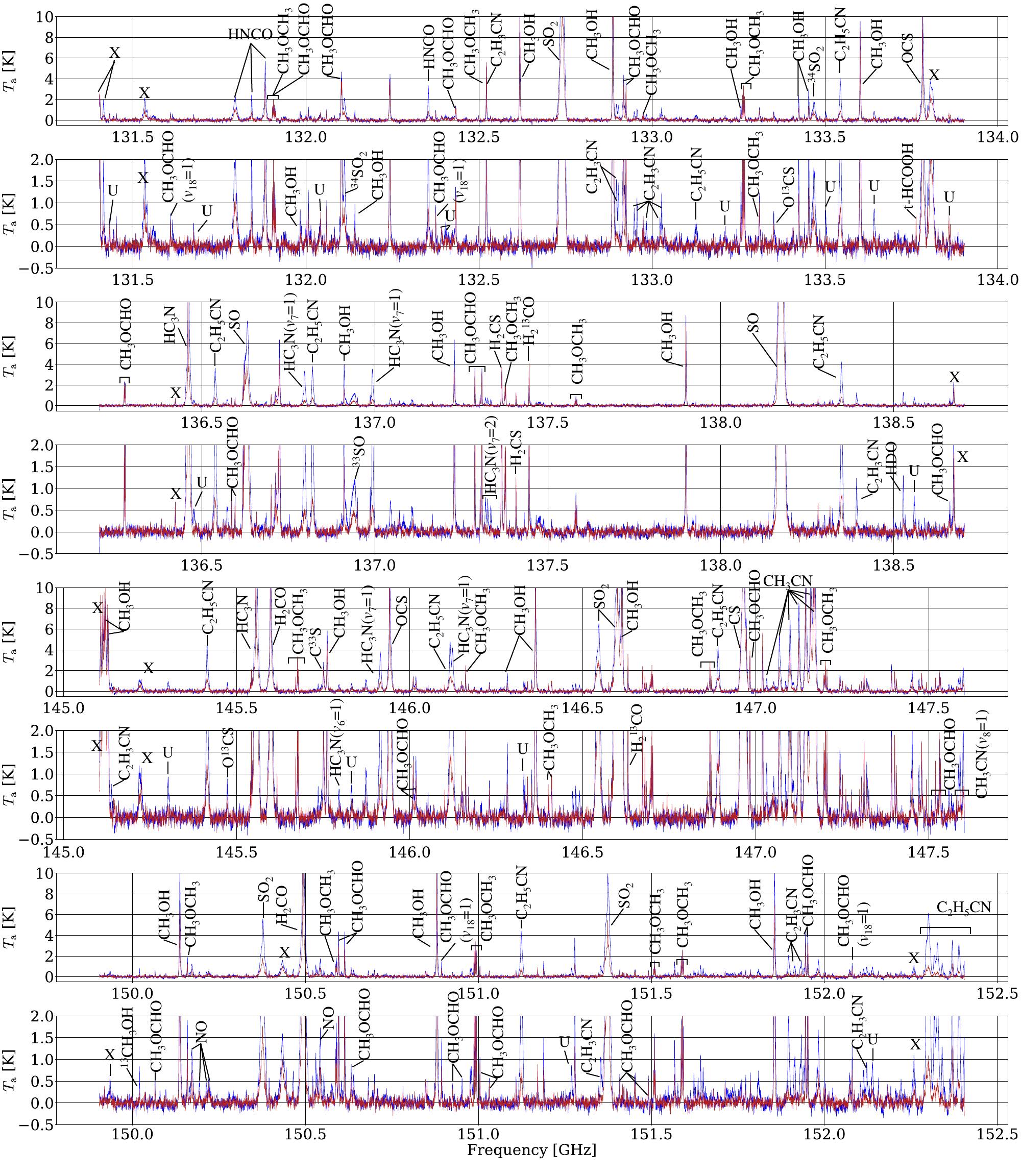}
    \caption {
        Spectra (in the $T_{\mathrm{a}}$ scale) taken toward the hot core (shown in blue) and compact ridge (shown in red) in Orion-KL.
        The spectra were obtained using the $5'\times5'$ OTF cube images by spatially averaging each circular region centered at the hot core and compact ridge, with a diameter of 11$''$--12$''$.
        They were also smoothed by binning 10 frequency channels from the original data, corresponding to a velocity resolution of, for example, 1.6\,km\,s$^{-1}$ at 140\,GHz.
        Two versions with different temperature scales (ranging from -0.5 to 2.0\,K and -0.5 to 10\,K) are shown for four different spectral windows.
        The line identifications are displayed in the spectra along with spurious signals (marked as ``$\times$'') caused by aliasing in the ADC, down-conversion, or leakage from the image sideband.
        Possible unidentified lines (U-lines) are also indicated (marked as ``U'').
    }
    \label{fig:lineid}
\end{figure*}

We obtained spectra on CASA by averaging within a circle with a diameter of 11$''$--12$''$, centered on both the hot core and compact ridge, using the OTF spectral fits made with channel binning.
Since the distance to Orion-KL is 418\,pc \citep{Kim2008-yg}, the spatial resolution of 11$''$--12$''$ corresponds to 0.02\,pc.
We covered a total of $\sim 16$\,GHz for the frequency range, from 127.9\,GHz to 152.4\,GHz.
The typical rms noise of the spectra is $\sim100$\,mK ($1\sigma$, 10\,ch binning) in $T_{\mathrm{a}}$ for the 2019 data.
Note that the on-source time for the spectra in the 2019 data is roughly 5\,s.
Nearly 400 lines were detected above the $3\sigma$ level from the $5'\times 5'$ OTF data.
Examples of the four different spectra from the $\rm 5'\times 5'$ OTF data covering 2.5\,GHz are shown in Figure~\ref{fig:lineid}.

The molecular line identification was carefully performed first by eye, rejecting spurious signals, contamination from corresponding other sidebands, aliasing signals due to A/D conversion in XFFTS, and contamination from out-band signals due to down-conversion.
Some results of the spectral line identifications are also shown in Figure~\ref{fig:lineid}.
We also used the \texttt{XCLASS} software \citep{xclass2017} on CASA.
The detailed results of the spectral line identifications and comparisons with previous studies using TRAO 14\,m, FCRAO 14\,m, and IRAM 30\,m telescopes \citep{Lee2001, Ziurys1993, Tercero2010} will be presented in the forthcoming paper by \citealt{Yonetsu2025}.
It is evident that the B4R system is very powerful for spectral scans.

By comparing the spectra between LSB and USB, we estimated the image rejection ratios of the receiver using strong lines such as SO ($T_{\mathrm{peak}}\sim45$\,K), H$_{2}$CO ($T_{\mathrm{peak}}\sim25.5$\,K).
We confirmed that the obtained values are in the range of 15 to 20\,dB and meet the specification.

    % !TEX root = ../main.tex
\section{Future Upgrade of the B4R System}
\label{sec:future-upgrade-of-the-b4r-system}

\subsection{Upgrade of the Receiver}
\label{sec:upgrade-of-the-receiver}

One of the important capabilities required for receivers on the LMT is the ability to be remotely tuned.
For such remote tuning of standard SIS receivers, optimizing LO power for each observing frequency setting (i.e., the corresponding LO frequency) is necessary to minimize the receiver noise temperature.
The LO power can be optimized using GPIB\footnote{General Purpose Interface Bus.}-controlled attenuators.
Another essential capability is receiver housekeeping, which involves monitoring the temperature of the 4\,K stage, vacuum gauge output, and other key parameters.
The systems required for the B4R were integrated into the current receiver system in 2024 May and are now ready for future use.

\subsection{Upgrade of Spectrometer}
\label{sec:upgrade-of-spectrometer}

The receiver output consists of four sets of 4--8\,GHz IF signals.
However, only four sets of 2.5\,GHz-wide IFs are currently processed using four XFFTS boards.
The cradle for the boards installed on the LMT can accommodate up to eight XFFTS boards.
Adding more boards and IF converters would enable processing of the full IF outputs.
Two additional XFFTS boards and the necessary IF components have already been prepared and will be integrated in the near future.

\subsection{Data-scientific Methods for Spectroscopic Observations}
\label{sec:data-scientific-methods-for-spectroscopic-observations}

The data-scientific method \citep{taniguchi2021}, i.e., low-rank and sparse decomposition using the GoDec algorithm \citep{Zhou2011}, applied to the PSW observations, is highly effective in achieving a higher S/N and eliminating systematic errors caused by atmospheric effects, such as spectral baseline ripples.
The proposed method demonstrated a 1.67-fold improvement in S/N in the integrated spectra compared to those obtained from the same dataset but analyzed using conventional methods, i.e., direct on-off source subtraction and polynomial baseline subtraction.
This improvement enables more efficient spectroscopy using the PSW mode, particularly for detecting faint molecular or atomic line emissions from high-$z$ objects.
Additionally, GoDec-based methods may also be applicable to the OTF mapping mode.
The detailed algorithm and analysis tool are currently under development.

The Frequency Modulation Local Oscillator \citep[FMLO;][]{taniguchi2020} is another data-scientific method for the PSW observations.
Since it does not require off-source measurements, the efficiency of on-source measurements is expected to improve by a factor of $\gtrsim2$, and software-based sideband separation is achieved.
The FMLO method was tested using the B4R on the LMT in 2018, demonstrating its basic functionality.
The system required for the FMLO method has been prepared\footnote{\url{https://github.com/b4r-dev/fmlolc-lmt}} and will be integrated into the B4R in the near future.

    % !TEX root = ../main.tex
\section{Summary}
\label{sec:summary}

We developed and installed the 2\,mm receiver system, B4R, on the 50\,m LMT telescope, located at an altitude of 4600\,m in Mexico.
The B4R receiver was developed based on the ALMA dual-polarization 2SB mixer technology and is equipped with an FFT digital spectrometer, XFFTS.
We tested three observing modes: the OTF mode, the PSW mode, and the spectral scan mode.

1. It was confirmed that the B4R receiver system installed on the LMT mostly meets the basic specifications in terms of receiver performance, such as receiver noise temperature and image rejection ratio.
Uranus was observed multiple times to estimate the aperture efficiency of the LMT.
The estimated values across the entire frequency range of the B4R (130--160\,GHz) were roughly consistent with those expected based on a surface accuracy of approximately 100\,$\mu$m and the receiver optics design.
Further investigation of the efficiencies will be required for a more accurate understanding of the telescope’s performance.

2. The OTF-mode observations were successfully performed toward OMC-1.
Thanks to the high sensitivity of the LMT, excellent sky conditions, and the wide frequency coverage of the XFFTS spectrometer while maintaining high spectral resolution, high-quality and high-sensitivity OTF maps were obtained for major lines such as CS and H$_{2}$CO , as well as for very weak lines such as C$^{33}$S and H35$\alpha$.

3. The PSW-mode observations were successfully performed toward lensed HyLIRGs at $z\sim2\text{--}4$.
It was confirmed that CO lines can be detected with a relatively flat baseline in the spectrum, covering 2.5\,GHz (corresponding to approximately 5000\,km\,s$^{-1}$ in velocity).
For very bright SMGs, CO or \textsc{[C\,i]} lines were detected with a 10-minute on-source integration.
Much fainter high-$z$ SMGs, where pointing sources are available within separations $\leq10\text{--}15$\,deg would be detectable in CO and \textsc{[C\,i]} lines with deeper integrations (i.e., repeating the PSW observations while inserting pointing calibrations at optimized time intervals).

4. The spectral scan capability was also tested and demonstrated using the OTF data, although no specifically designed spectral scan observation has been conducted in the science demonstration.

5. Currently, the B4R on the LMT, which has two polarizations and four available XFFTS boards, is operational for general observing with the remote tuning system.
However, a few remaining issues need to be resolved as soon as possible, such as upgrading the spectrometer to six XFFTS boards (with a future expansion to eight full boards).
We have demonstrated that the B4R on the LMT enables the most sensitive single-dish spectroscopic observations at 2\,mm.

    % !TEX root = ../main.tex
\begin{acknowledgments}
    This work was financially supported by JSPS KAKENHI grant Nos., JP15H02073, 17H06130 (R.K., Y.T., K.K.), JP19K14754 (T.T.), 22H04939 (T.S., K.T., A.T., Y.T., K.K.), JP19K21885 and JP25K01060 (H.M.), JP20K14523, JP21H01142, JP24K17096, JP24H00252 (K.T.), 22J21948, 22KJ1598 (M.H.), JP22J22889, and JP22KJ2625 (T.Y.).
    T.T. is supported by MEXT Leading Initiative for Excellent Young Researchers (grant No. JPMXS0320200188).
    This paper makes use of data obtained with the Large Millimeter Telescope Alfonso Serrano (LMT) in Mexico.
    The LMT project is a joint effort between the Instituto Nacional de Astr\'{o}fisica, \'{O}ptica, y Electr\'{o}nica (INAOE) and the University of Massachusetts at Amherst (UMASS).
    We also appreciate the support of the LMT technical staff and support scientists during the commissioning campaign of the B4R.
    Data analysis was partially carried out on the Multi-wavelength Data Analysis System, operated by the Astronomy Data Center (ADC), National Astronomical Observatory of Japan.
\end{acknowledgments}

\facility{
    LMT (B4R)
}

\software{
    Astropy \citep{Astropy2013,Astropy2018},
    CASA \citep{CASA2022},
    fmflow \citep{Taniguchi2022},
    NumPy \citep{Harris2020},
    matplotlib \citep{Hunter2007},
    scikit-learn \citep{Pedregosa2011},
    SciPy \citep{Virtanen2020},
    pandas \citep{Pandas2020,McKinney2010},
    xarray \citep{Hoyer2017},
    XCLASS \citep{xclass2017}.
}

    % !TEX root = ../main.tex
\appendix

\section{Emission-line Spectra Detected in Other High Redshift HyLIRGs}
\label{sec:spectra-of-other-emission-line-detected-in-high-z-smgs}

\begin{figure*}[htbp]
    \centering
    \includegraphics[width=\linewidth]{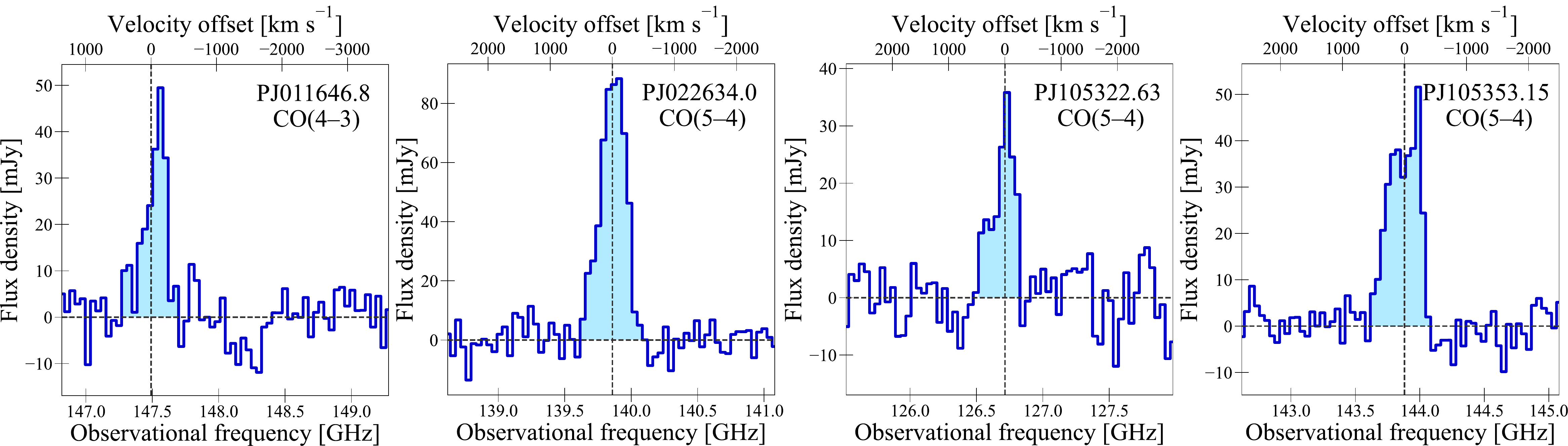}
    \label{fig:high-z-observation_appendix}
    \caption{
        Redshifted CO spectra taken toward four other Planck-selected, strongly lensed HyLIRGs at $z=2\text{--}3.5$.
        The spectroscopic redshifts of these sources were previously identified via CO emission-line detections (\citealt{Harrington2016, Harrington2021, Canameras2018}).
        In each panel, $v=0$\,km\,s$^{-1}$ corresponded to the line redshift obtained from our observations (see Table~\ref{tab:high-z-observation_appendix}).
    }
    \label{fig:high-z-spec_appendix}
\end{figure*}

Figure~\ref{fig:high-z-spec_appendix} shows all the other spectra for emission-line-detected high-$z$ lensed HyLIRGs, except for PJ020941.3, which is presented in Section~\ref{sec:psw-observations-of-high-z-smgs}.
The spectroscopic redshifts of these sources were also identified via CO lines in previous studies (\citealt{Harrington2016, Harrington2021, Canameras2018}).
Table~\ref{tab:high-z-observation_appendix} summarizes the targets and our observations.

\begin{table*}[htbp]
    \centering
    \caption{Observation logs of high-$z$ targets obtained with B4R on the LMT}
    \label{tab:high-z-observation_appendix}
    \scalebox{0.7}{
        \begin{tabular}{lllll}
            \hline
            Target name & PJ011646.8 & PJ022634.0 & PJ105322.63 & PJ105353.15\\
            Target coordinates (R.A., Dec.) (J2000) & ($01^\mathrm{h}16^\mathrm{m}48.8^\mathrm{s}$, $-24^\circ37^\prime02''$) & ($02^\mathrm{h}26^\mathrm{m}34.0^\mathrm{s}$, $+23^\circ45^\prime28''$) & ($10^\mathrm{h}53^\mathrm{m}22.63^\mathrm{s}$, $+60^\circ51^\prime47.1''$) & ($10^\mathrm{h}53^\mathrm{m}53.15^\mathrm{s}$, $+05^\circ56^\prime18.8''$)\\
            Target line & CO\,$J=4\text{--}3$ & CO\,$J=5\text{--}4$ & CO\,$J=5\text{--}4$ & CO\,$J=5\text{--}4$\\
            \hline\hline
            Observation date & 2019 Nov 28 & 2019 Nov 27 & 2019 Nov 28 & 2019 Nov 26\\
            The total number of scans & 2 & 1 & 1 & 2\\
            $T_\mathrm{atm}~\mathrm{[K]}$ & 275.2--275.4 & 277.1 & 274.8 & 276.6--276.7\\
            Opacity at $220~\mathrm{GHz}$ & 0.08 & 0.10 & 0.10 & 0.12\\
            System noise temperature [K] (min, max) & (134, 210) & 112 & (124, 129) & (94, 98)\\
            \hline
            First LO frequency [GHz] & 141.2 & 146.7 & 133.6 & 137.0\\
            XFFTS frequency range [GHz] & 146.8--149.3 & 138.6--141.1 & 125.5--128.0 & 142.6--145.1\\
            512-ch binned channel width [GHz] & 0.04 (79\,km\,s$^{-1}$) & 0.04 (84\,km\,s$^{-1}$) & 0.04 (92\,km\,s$^{-1}$) & 0.04 (81\,km\,s$^{-1}$)\\
            \hline
            Integration time at on-source position [s] & 600 & 300 & 300 & 300\\
            Integration time at off-source position [s] & 600 & 300 & 300 & 300\\
            Baseline subtracted function & linear $+$ sine curve & linear & linear & linear\\
            \hline
            Line redshift & $2.1255\pm0.0001$ & $3.1204\pm0.0001$ & $3.5486\pm0.0001$ & $3.0054\pm0.0001$\\
            Noise level [mJy] & 5.07 & 5.03 & 4.92 & 3.71\\
            \hline
        \end{tabular}
    }
\end{table*}

    \bibliography{references}
    \bibliographystyle{aasjournal}
\end{document}